\documentclass[11pt,a4paper]{article}

\usepackage[utf8]{inputenc}
\usepackage[T1]{fontenc}
\IfFileExists{lmodern.sty}{\usepackage{lmodern}}{}
\IfFileExists{microtype.sty}{\usepackage[expansion=false,protrusion=true]{microtype}}{}

\usepackage{amsmath,amssymb,amsthm}
\usepackage{graphicx}
\usepackage{booktabs}
\usepackage{float}
\usepackage{caption}
\usepackage[margin=2.4cm]{geometry}
\usepackage{hyperref}
\usepackage{cite}
\usepackage{xcolor}

\hypersetup{
  colorlinks=true,
  linkcolor=blue!55!black,
  citecolor=blue!55!black,
  urlcolor=blue!55!black
}

\graphicspath{{./figures/}{./}}
\newcommand{\ups}[1]{\ensuremath{\Upsilon(#1\mathrm{S})}}
\newcommand{\psipr}{\ensuremath{\psi(2\mathrm{S})}}
\newcommand{\Jpsi}{\ensuremath{J/\psi}}
\newcommand{\Ntrack}{\ensuremath{N_{\mathrm{track}}}}
\newcommand{\Ncone}{\ensuremath{N^{\Delta R}_{\mathrm{track}}}}
\newcommand{\Nphi}{\ensuremath{N^{\Delta\phi}_{\mathrm{track}}}}
\newcommand{\ST}{\ensuremath{S_{T}}}
\newcommand{\Rtwo}{\ensuremath{R_{21}}}
\newcommand{\Rthree}{\ensuremath{R_{31}}}
\newcommand{\Rthirtytwo}{\ensuremath{R_{32}}}
\newcommand{\dNdeta}{\ensuremath{dN_{\mathrm{ch}}/d\eta}}
\newcommand{\pT}{\ensuremath{p_{T}}}
\newcommand{\DR}{\ensuremath{\Delta R}}
\newcommand{\Dphi}{\ensuremath{\Delta\phi}}
\newcommand{\bbbar}{\ensuremath{b\bar b}}
\newcommand{\ccbar}{\ensuremath{c\bar c}}

\newcommand{\sqrts}{\ensuremath{\sqrt{s}}}

\begin{document}

\title{\textbf{ Sequential Y(nS) suppression in high-multiplicity pp collisions:
 the experimental case for an early, globally correlated medium}}
\author{Renato Campanini\\[3pt]
\small Dipartimento di Fisica e Astronomia, Universit\`a di Bologna\\
\small and INFN, Sezione di Bologna, Italy}
\date{}
\maketitle

\begin{abstract}
The multiplicity-dependent suppression of $\Upsilon(n\mathrm{S})$ states
measured by CMS in $pp$ at $\sqrts=7$\,TeV~\cite{CMS2020}, and of
$\psipr/\Jpsi$ measured by LHCb at $\sqrts=13$\,TeV~\cite{LHCb2024},
is subjected to four multi-differential tests:
\emph{cone isolation}, \emph{azimuthal sectors}, \emph{transverse
sphericity}, and \emph{prompt vs.\ non-prompt}.
Cone and sphericity close a \emph{scissors constraint}: the local
reading of the Comover Interaction Model is in tension with the cone
data, its global reading with the sphericity data.
The non-prompt flatness forces the mechanism to act at early proper
times.
None of the considered hadronic or string-based frameworks --- CIM
local or global, PYTHIA 8 MPI~\cite{Sjostrand2015}, rope
hadronisation~\cite{Bierlich2015}, CGC~\cite{Ma2015}, Trainor
TCM~\cite{Trainor2008} --- naturally satisfies the four constraints
simultaneously in its published form.
The surviving class is an early, globally correlated medium consistent
with partonic degrees of freedom, co-occurring with the ALICE
strangeness enhancement~\cite{ALICE_SE}, the long-range
ridge~\cite{CMS_ridge}, and below the threshold of the partonic
baryon--meson $v_2$~\cite{ALICE_v2}, in a density window compatible
with the Campanini \& Ferri equation of state~\cite{Campanini2011}.
\end{abstract}

\section{Introduction}
The question at the heart of this paper can be stated in one sentence:
is the suppression of the excited $\Upsilon$ states observed in
high-multiplicity proton--proton collisions driven by ordinary hadronic
final-state scattering, or is it telling us that the smallest collision
system at the LHC already produces a transient, partonic-like medium?
For decades, the progressive melting of quarkonium bound states has been
read as the canonical thermometer of colour deconfinement in
heavy-ion physics~\cite{MatsuiSatz1986,Satz2006}: the larger and more
loosely bound a state, the lower the temperature at which Debye
screening dissolves it.
Applied to bottomonium, this picture predicts a clean hierarchy in
which $\ups{3}$, the fragile partner with a binding energy of only
$\sim 200$\,MeV, melts first; $\ups{2}$ next; and $\ups{1}$, tightly
bound by more than an order of magnitude, last.
The same hierarchy has now been measured by
CMS at $\sqrts = 7$\,TeV~\cite{CMS2020} and by
LHCb at $\sqrts = 13$\,TeV~\cite{LHCb2024,LHCb2025} in plain $pp$
collisions --- a system where, by textbook reasoning, no collective
medium should form at all.
The \emph{fact} of the hierarchy is therefore not in question; what is
in question is its \emph{origin}.

Three explanations compete for the role.
The first and most conservative is the Comover Interaction Model, in
which the quarkonium is dissociated by inelastic scattering off the
soft hadrons produced in the same rapidity
region~\cite{Armesto1998,Gavin1990,Ferreiro2018}.
The same framework has been extended beyond conventional states to the
exotic $X(3872)$ in high-multiplicity $pp$~\cite{Esposito2021}, where
it plays a central role in constraining the hadronic molecule versus
compact tetraquark interpretation.
It is natural: the denser the event, the more scatterers; the more
scatterers, the more dissociation; the larger the state, the larger
the cross section.
The second ascribes the effect to initial-state or string-topology
phenomena that modify the production step itself --- Colour Glass
Condensate at low $x$~\cite{Ma2015}, or the enhanced effective string
tension of rope hadronisation~\cite{Bierlich2015} when overlapping
strings from many multi-parton interactions rearrange.
The third invokes a genuine, short-lived partonic droplet, with colour
screening acting on the pre-resonance $\bbbar$ pair before the
$\Upsilon$ state has even had time to form.
The three pictures are not just different in name: they make different
predictions for observables beyond the inclusive multiplicity trend,
and this is what allows the data to discriminate between them.

The key is that both CMS and LHCb did not stop at measuring the ratios
against the global multiplicity.
They also resolved the events by the local track density around the
$\Upsilon$ (cone isolation), by the direction of the extra activity
with respect to the $\Upsilon$ (azimuthal sectors), by the global shape
of the event (transverse sphericity), and --- crucially --- by the
production mechanism itself, separating quarkonia born at the primary
vertex from those produced in the displaced decays of $b$ hadrons.
Each of these cuts is a test with a definite prediction.
Local hadronic scattering must depend on local hadron density.
A mechanism driven only by the total multiplicity cannot depend on the
shape of the event at fixed multiplicity.
A mechanism acting at late proper times must affect displaced quarkonia
just as it affects prompt ones.
The surprise, and the content of this paper, is that when one imposes
all the cuts simultaneously \emph{every} hadronic or string-based
candidate fails at least one of them, while a short-lived partonic
medium satisfies them jointly.

What makes this conclusion more than an isolated quarkonium argument
is the context in which it sits.
The onset of $\Upsilon$ suppression in $pp$ lines up, in the same
multiplicity range, with the strangeness enhancement seen by
ALICE~\cite{ALICE_SE} --- the oldest and cleanest QGP signal known from
heavy-ion physics --- and with the long-range two-particle ridge that
developed into a unique feature of high-multiplicity $pp$ data at the
LHC~\cite{CMS_ridge}.
At still higher multiplicities, ALICE has recently reported the
baryon--meson $v_2$ grouping~\cite{ALICE_v2}, the mass-ordering switch
that is the fingerprint of partonic, rather than hadronic,
collectivity.
Independently, the equation-of-state analysis of Campanini \&
Ferri~\cite{Campanini2011} places a soft-to-hard transition band
in $pp$ and $p\bar p$ data across ISR-to-LHC energies precisely where
all these phenomena switch on.
The absence of jet quenching in the same high-multiplicity $pp$
collisions is not a contradiction: path-length scaling alone suppresses
any quenching signal below current detector sensitivity for a droplet
of $\mathcal{O}(1)$\,fm.
Seen together, these are a convergent set of experimental observations
that single out a specific class of mechanisms.

\section{Notation and observables}
\label{sec:notation}
Throughout the paper we use the following symbols and conventions.
\begin{description}
  \item[$\Ntrack$] charged-track multiplicity, CMS acceptance
  $|\eta|<2.4$, $\pT>0.4$\,GeV/$c$.
  The relation $\Ntrack\simeq 2.83\,\dNdeta$ simply reflects the
  integration of $\dNdeta$ over the $|\eta|<2.4$ window after the
  reconstruction and $\pT$-selection efficiencies have been folded in;
  the explicit conversion table
  ($\Ntrack$ measured $\leftrightarrow$ $\Ntrack$ true
  $\leftrightarrow$ $\dNdeta$ for $\pT>0$) is provided by CMS in
  Ref.~\cite{CMS2020}.
  \item[$\Rtwo,\Rthree$] production ratios normalised to the ground
  state, $R_{n1}=Y(\ups{n})/Y(\ups{1})$, $n=2,3$, for prompt yields $Y$.
  \item[$\Rthirtytwo$] the differential ratio
  $\Rthirtytwo=Y(\ups{3})/Y(\ups{2})=\Rthree/\Rtwo$, which cancels
  common-mode systematics (muon trigger, reconstruction, correlated
  feed-down) and isolates the sequential hierarchy.
  \item[$\Ncone$] tracks with $\pT>0.4$\,GeV/$c$ inside a cone
  $\DR<0.5$ around the $\Upsilon$ flight axis, where
  $\DR\equiv\sqrt{(\Delta\eta)^2+(\Dphi)^2}$.
  \item[$\Nphi$] sector multiplicity as a function of azimuthal angle
  $\Dphi$ between a track and the $\Upsilon$:
  forward $|\Dphi|<\pi/3$,
  transverse $\pi/3<|\Dphi|<2\pi/3$,
  backward $|\Dphi|>2\pi/3$.
  \item[$\ST$] transverse sphericity,
  $\ST=2\lambda_2/(\lambda_1+\lambda_2)$, from the two eigenvalues
  of the transverse momentum tensor;
  $\ST\to 0$ jet-like, $\ST\to 1$ isotropic.
  \item[$\pT^\Upsilon$] transverse momentum of the reconstructed $\Upsilon$.
\end{description}

The paper is organised as follows.
Sec.~\ref{sec:basic} presents the multiplicity dependence of $\Rtwo$ and
$\Rthree$ inclusive and at high $\pT$.
Sec.~\ref{sec:cone} uses the cone-isolation test to falsify local
hadronic mechanisms.
Sec.~\ref{sec:azimuth} uses the azimuthal-sector data to remove any
residual near-side scenario.
Sec.~\ref{sec:sphericity} uses the transverse sphericity to falsify
every mechanism that is a function of $\Ntrack$ alone.
Sec.~\ref{sec:nonprompt} introduces the LHCb non-prompt $\psipr/\Jpsi$
temporal constraint.
Sec.~\ref{sec:models} gathers the verdicts on the hadronic and
string-based frameworks.
Sec.~\ref{sec:context} connects the onset to strangeness enhancement,
to the ridge, and to the ALICE partonic-flow observation.
Sec.~\ref{sec:eos} notes the coincidence of the onset multiplicity
window with the transition band of Campanini \& Ferri.
Sec.~\ref{sec:nojet} explains why the absence of jet quenching in $pp$
is \emph{compatible} with a partonic droplet.

\section{Suppression in multiplicity: inclusive and high-$p_T$}
\label{sec:basic}
\begin{figure}[H]
\centering
\includegraphics[width=\textwidth]{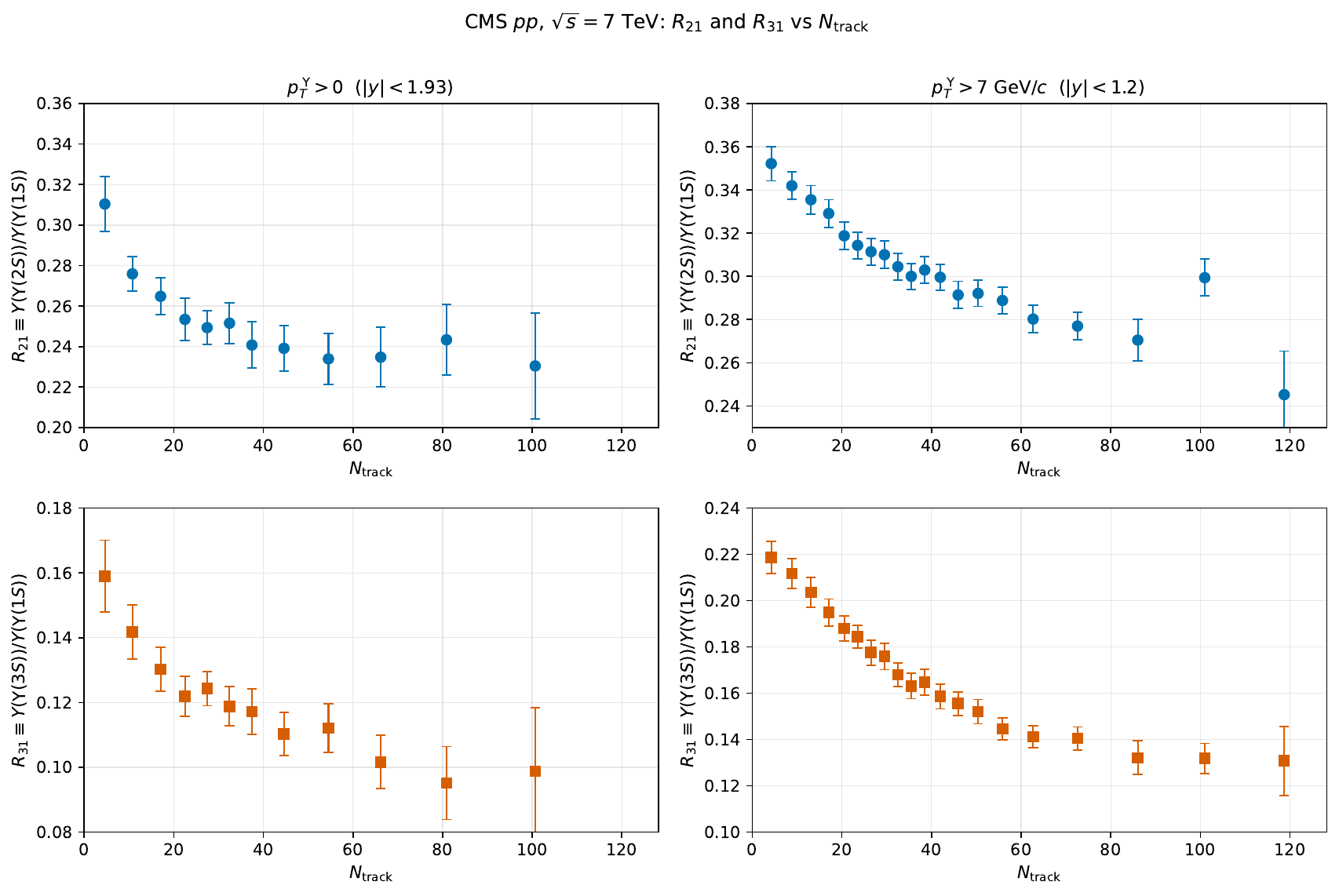}
\caption{CMS $pp$ data at $\sqrts=7$\,TeV
(own plot from HEPData of Ref.~\cite{CMS2020}).
Top row: $\Rtwo=Y(\ups{2})/Y(\ups{1})$.
Bottom row: $\Rthree=Y(\ups{3})/Y(\ups{1})$.
Left column: inclusive $\pT^\Upsilon>0$, $|y|<1.93$.
Right column: $\pT^\Upsilon>7$\,GeV/$c$, $|y|<1.2$.
In all four panels the ratio decreases monotonically with $\Ntrack$;
the high-$\pT$ sample covers a wider multiplicity range and shows a
cleaner monotonic trend.
$\ups{3}$ is always more suppressed than $\ups{2}$ (sequential
suppression).}
\label{fig:R_vs_N}
\end{figure}

The basic experimental fact is the monotonic decrease of both $\Rtwo$
and $\Rthree$ with $\Ntrack$ in both event classes measured by CMS
(Fig.~\ref{fig:R_vs_N}).
Two features matter for what follows.

First, the suppression is \emph{monotonic} in $\Ntrack$ for both
$\Rtwo$ and $\Rthree$, and the relative suppression of $\Rthree$ is
stronger than that of $\Rtwo$ --- the expected sequential hierarchy by
state size.

Second, at the same $\Ntrack$ the high-$\pT$ sample lies
\emph{systematically above} the inclusive one.
Reading Fig.~\ref{fig:R_vs_N} at corresponding multiplicity bins, the
high-$\pT$ ratios are
$\Rtwo(\pT>7)/\Rtwo(\pT>0)\simeq 1.2$ and
$\Rthree(\pT>7)/\Rthree(\pT>0)\simeq 1.4$ across the multiplicity range
accessible to both samples.
In other words, requiring a fast $\Upsilon$ reduces the amount of
suppression, and reduces it more strongly for the more loosely bound
state.
This is the qualitative signature one expects from a kinematic
\emph{escape}: a fast quarkonium crosses the dense region in a time
$\tau_{\mathrm{cross}}\sim R/(\beta\gamma)$ that shortens with
$\pT^\Upsilon$, so a larger fraction of the population escapes
intact, and the escape fraction is naturally higher for the more
fragile states.

\begin{figure}[H]
\centering
\includegraphics[width=\textwidth]{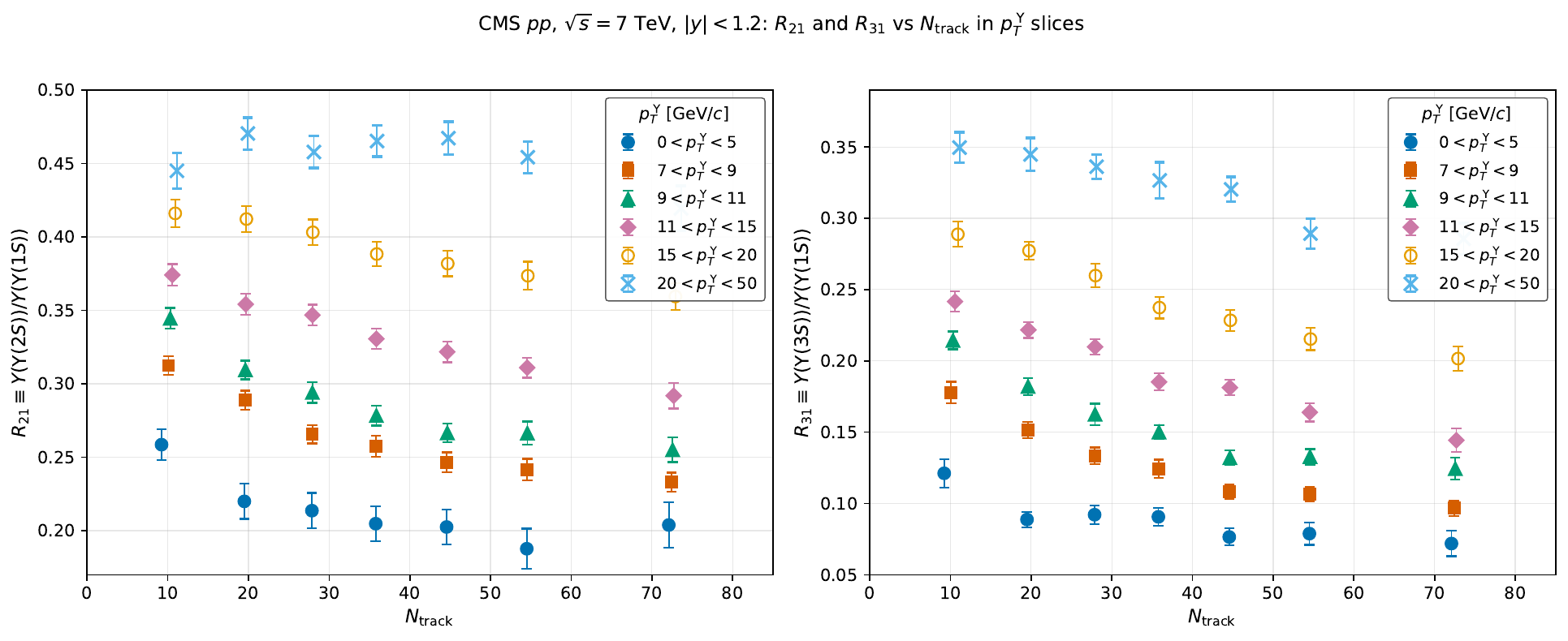}
\caption{CMS $pp$ data at $\sqrts=7$\,TeV~\cite{CMS2020}:
$\Rtwo$ (left) and $\Rthree$ (right) as a function of $\Ntrack$ in six
exclusive $p_T^\Upsilon$ slices,
$0<p_T^\Upsilon<5$, $7<p_T^\Upsilon<9$, $9<p_T^\Upsilon<11$,
$11<p_T^\Upsilon<15$, $15<p_T^\Upsilon<20$, $20<p_T^\Upsilon<50$\,GeV/$c$
(from dark to light, all for $|y|<1.2$).
Own plot from the public HEPData tables of Ref.~\cite{CMS2020}.
At fixed $\Ntrack$ the slices are clearly ordered by $p_T^\Upsilon$:
higher momentum gives systematically higher $R_{n1}$.
The $\Ntrack$-dependence itself evolves with $p_T^\Upsilon$: the low-$\pT$
slices show the strongest decrease with multiplicity, while the
$20<p_T^\Upsilon<50$ slice is compatible with flat behaviour.
The effect is more pronounced for $\Rthree$ than for $\Rtwo$.}
\label{fig:ptslices}
\end{figure}

A fully differential picture of the effect is provided by the CMS
$p_T^\Upsilon$-sliced measurement~\cite{CMS2020}, which we replot in
Fig.~\ref{fig:ptslices}.
Two observations follow directly from the figure:
(i) at fixed $\Ntrack$ the six slices are ordered from bottom to top by
increasing $p_T^\Upsilon$ for both $\Rtwo$ and $\Rthree$, so the
$\pT$-dependent survival is not an artefact of inclusive-sample
averaging;
(ii) the slope of $R_{n1}$ versus $\Ntrack$ weakens monotonically as
$p_T^\Upsilon$ increases, to the point that the highest-$\pT$ slice is
compatible with no $\Ntrack$-dependence within uncertainty, as one
expects if the $\Upsilon$ crossing time becomes much shorter than the
dissociation timescale.
This is not a measurement of a specific dissociation cross section,
but it is the differential evidence, in the CMS data themselves, for a
$\pT^\Upsilon$-dependent survival mechanism.
The LHCb $\psipr/\Jpsi$ analysis of Sec.~\ref{sec:nonprompt} provides
a consistent $\pT$-resolved view on charmonium.

\begin{figure}[H]
\centering
\includegraphics[width=\textwidth]{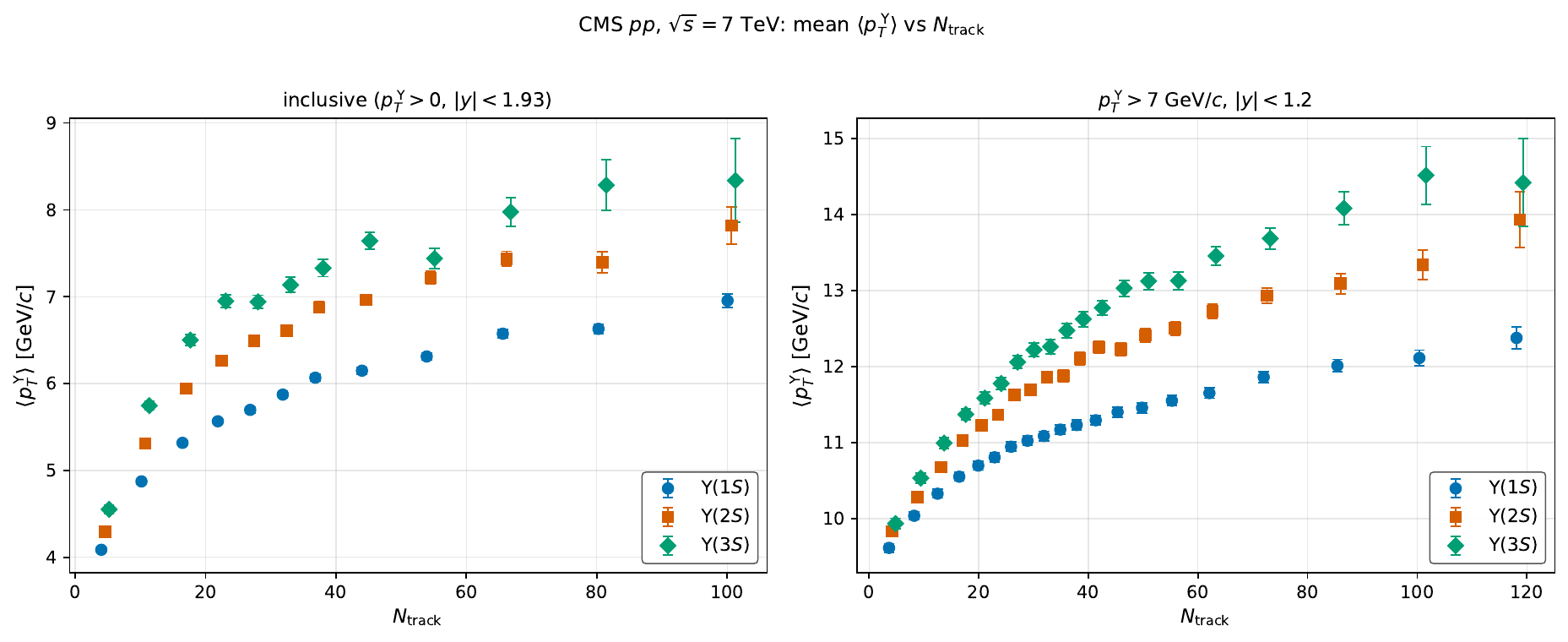}
\caption{Mean transverse momentum $\langle p_T^\Upsilon\rangle$ of the
three $\Upsilon(n\mathrm{S})$ states as a function of $\Ntrack$,
separately for the inclusive sample ($\pT^\Upsilon>0$, $|y|<1.93$, left)
and the high-$\pT$ sample ($\pT^\Upsilon>7$\,GeV/$c$, $|y|<1.2$, right).
Own plot from the public HEPData tables of Ref.~\cite{CMS2020}.
At low $\Ntrack$ the three means coincide, confirming that the
three states share the same production kinematics.
As $\Ntrack$ increases, the means split into the hierarchy
$\langle p_T\rangle(\ups{3}) > \langle p_T\rangle(\ups{2}) >
\langle p_T\rangle(\ups{1})$.
The same ordering is present in both kinematic samples; the absolute
scale is set by the $\pT$ cut.}
\label{fig:meanpt}
\end{figure}

Third, the mean transverse momentum of the three $\Upsilon$ states as a
function of $\Ntrack$ carries the same message in a complementary,
\emph{inclusive-yield-weighted} form (Fig.~\ref{fig:meanpt}).
At the lowest accessible multiplicity the three distributions have
essentially the same mean $\pT^\Upsilon$: the three states are produced
with the same kinematics.
As $\Ntrack$ grows, the three means fan out and order themselves into
$\langle p_T\rangle(\ups{3}) > \langle p_T\rangle(\ups{2}) >
\langle p_T\rangle(\ups{1})$.
The ordering by state size is the same in the inclusive and in the
high-$\pT$ sample; only the absolute scale changes with the kinematic
cut.
This is consistent with a picture in which a dissociative mechanism
depletes each state preferentially at low $\pT^\Upsilon$, so the
surviving $\Upsilon$ population gets progressively ``pushed'' to
higher $\pT$; and since the more loosely bound states are depleted
more strongly, their surviving mean shifts upward more.
We stress that Fig.~\ref{fig:meanpt} is a shape statement about the
surviving population --- not an independent dynamical measurement ---
and that alternative production-level explanations for the same
ordering cannot be ruled out on inclusive yields alone.
Taken together with Fig.~\ref{fig:R_vs_N}, however, the two
independent readings of the data point in the same direction: a
$\pT^\Upsilon$-dependent survival, stronger for the more loosely
bound states.
These two features set the stage but do not discriminate between the
candidate mechanisms.
The discrimination comes from the differential tests of the following
sections.

\section{Cone isolation constrains local hadronic CIM}
\label{sec:cone}
\paragraph{The test.}
CMS~\cite{CMS2020} splits events at fixed $\Ntrack$ into two extreme
classes according to the number of tracks inside a cone $\DR<0.5$
around the $\Upsilon$:
\begin{itemize}
\item \emph{empty cone}, $\Ncone=0$, no charged track (above
$\pT>0.4$\,GeV/$c$) collinear with the $\Upsilon$;
\item \emph{dense cone}, $\Ncone>2$, at least three such tracks.
\end{itemize}
Any mechanism in which the $\Upsilon$ is dissociated by inelastic
scattering with \emph{nearby} hadrons must produce different suppression
in the two classes: more local scatterers $\Rightarrow$ stronger
suppression.
The sign of the expected effect is unambiguous.

\paragraph{The data.}
\begin{figure}[H]
\centering
\includegraphics[width=\textwidth]{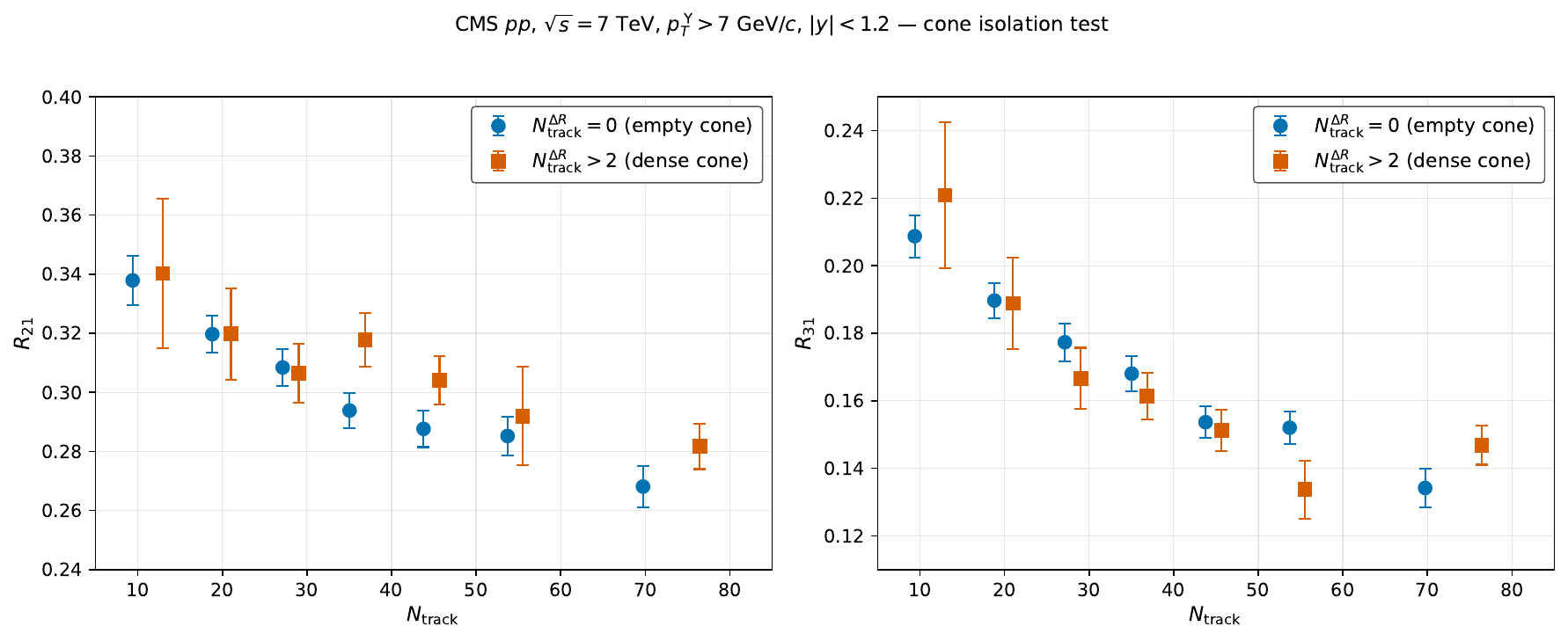}
\caption{Cone-isolation test at $\sqrts=7$\,TeV.
Left: $\Rtwo$.
Right: $\Rthree$.
Blue circles: empty cone ($\Ncone=0$).
Red diamonds: dense cone ($\Ncone>2$).
Own plot from HEPData of Ref.~\cite{CMS2020}, $\pT^\Upsilon>7$\,GeV/$c$,
$|y|<1.2$.
The two classes are statistically indistinguishable across
$10\lesssim\Ntrack\lesssim 80$; any local hadronic mechanism predicts a
clear separation that is not seen in the data.}
\label{fig:cone}
\end{figure}
At $\Ntrack\simeq 70$ the empty-cone suppression is $-20.7\%$ relative
to the lowest-$\Ntrack$ bin, the dense-cone value is $-17.2\%$, and the
two are compatible within uncertainty.

\paragraph{Consequence for hadronic CIM (local reading).}
Any mechanism in which the quarkonium is dissociated by inelastic
scattering with \emph{nearby} produced hadrons requires the dissociation
probability to grow monotonically with the local hadron density
around the $\Upsilon$~\cite{Armesto1998,Gavin1990,Ferreiro2018}.
This expectation is model-independent: the word ``comover'' itself
implies that proximity matters, and more comovers in the cone must
produce more suppression, irrespective of the precise form of the
dissociation cross section or the binding potential.
The sign of the effect is unambiguous.

Keeping the three levels separate:
\emph{Data}: the empty-cone and dense-cone classes are
statistically indistinguishable within uncertainty across the full
$\Ntrack$ range accessible to both.
\emph{Experimental constraint}: the suppression does not scale with the
local track density around the $\Upsilon$; whatever variable drives the
suppression, it is not the number of hadrons collinear with the
quarkonium.
\emph{Interpretation}: in its local-density form, any comover-like
model is in direct tension with the cone data.
The framework can of course be preserved by promoting the comover
density to a global quantity; Sec.~\ref{sec:sphericity} shows that
this move is itself in tension with the sphericity-resolved data.
This closes the first door.

\section{Azimuthal independence removes the ``near-side'' escape}
\label{sec:azimuth}
The cone test removes local scattering in a \emph{forward} cone around
the $\Upsilon$.
A local mechanism could, however, be imagined where tracks
\emph{collinear} with the $\Upsilon$ drive the suppression while tracks
in other directions do not.
CMS resolves this by measuring the suppression as a function of the
multiplicity restricted to three non-overlapping azimuthal sectors
(forward, transverse, backward) relative to the $\Upsilon$ flight
direction.
\begin{figure}[H]
\centering
\includegraphics[width=\textwidth]{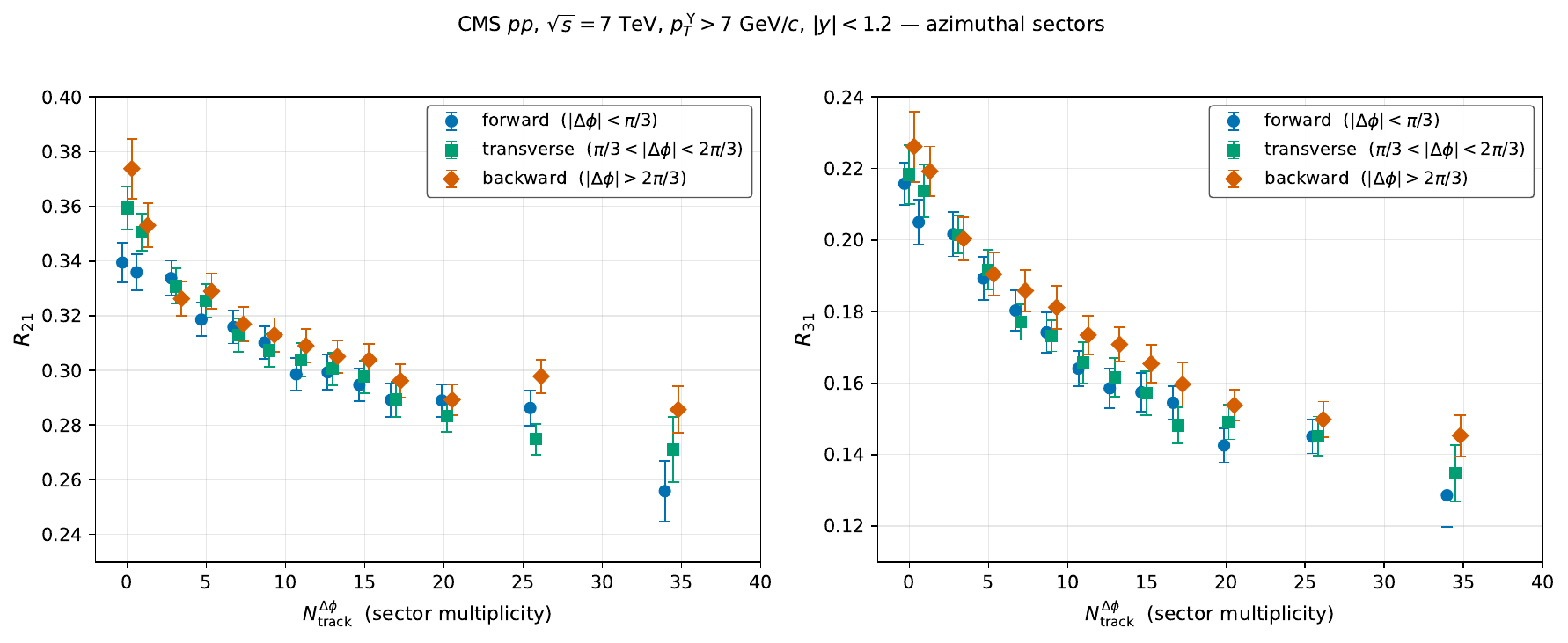}
\caption{Suppression ratios as a function of sector multiplicity
$\Nphi$ in three azimuthal bins relative to the $\Upsilon$:
forward ($|\Dphi|<\pi/3$), transverse ($\pi/3<|\Dphi|<2\pi/3$),
backward ($|\Dphi|>2\pi/3$).
Own plot from HEPData of Ref.~\cite{CMS2020},
$\pT^\Upsilon>7$\,GeV/$c$, $|y|<1.2$.
Left: $\Rtwo$.
Right: $\Rthree$.
All three sector datasets are statistically consistent within
uncertainty: the suppression scales with the sector multiplicity at the
same rate in every direction.
Tracks on the \emph{opposite} side of the $\Upsilon$ (backward) drive
the suppression as efficiently as tracks collinear with it (forward).
No near-side hierarchy survives this test.}
\label{fig:azimuth}
\end{figure}
The backward result is particularly instructive: backward tracks are
geometrically far from the $\Upsilon$, so they cannot represent a local
scattering environment.
Yet the most compact and complete statement is provided by the
\emph{transverse} sector ($\pi/3<|\Dphi|<2\pi/3$): tracks at 90°
to the $\Upsilon$ flight direction are neither co-moving nor recoiling
against it, and have no preferred geometric or kinematic relationship
to the $\Upsilon$ trajectory.
Their multiplicity nonetheless correlates with the suppression at the
same rate as forward tracks.
The three-fold azimuthal symmetry --- forward, transverse, and backward
sectors all yielding statistically identical suppression slopes ---
is therefore a stronger statement than a forward--backward comparison
alone: it closes not only the near-side escape but also any model
based on recoil, deflection, or directional energy flow.
What drives the suppression is a
\emph{global} activity variable, of which all three sector multiplicities
are equivalent proxies.
Azimuthal independence is therefore a \emph{necessary} but not
\emph{sufficient} element: it rules out proximity-based and
direction-based mechanisms, but does not by itself rule out a mechanism
in which $\Ntrack$ itself is the relevant variable.
The next step discriminates this remaining possibility.

\section{Sphericity constrains every $N_{\text{track}}$-only mechanism}
\label{sec:sphericity}
\paragraph{The test.}
At fixed $\Ntrack$, the transverse sphericity $\ST$ separates events
with the same global multiplicity into radically different
\emph{topologies}:
\begin{itemize}
\item $\ST<0.55$: jet-like events, dominated by a hard process with a
strong back-to-back axis;
\item $\ST>0.85$: isotropic events, dominated by many soft MPI
distributed over azimuth.
\end{itemize}
Any mechanism that parametrises the dissociation probability as a
function of $\Ntrack$ alone \emph{must} give the same suppression in
the two classes.
This includes: the hadronic global CIM
($\rho_{\mathrm{com}}\propto\Ntrack$), the Colour Glass Condensate
(initial-state $Q_s^2$ set by the overall collision density), and pure
PYTHIA MPI (no interaction with the produced $\Upsilon$).

\paragraph{The data.}
\begin{figure}[H]
\centering
\includegraphics[width=\textwidth]{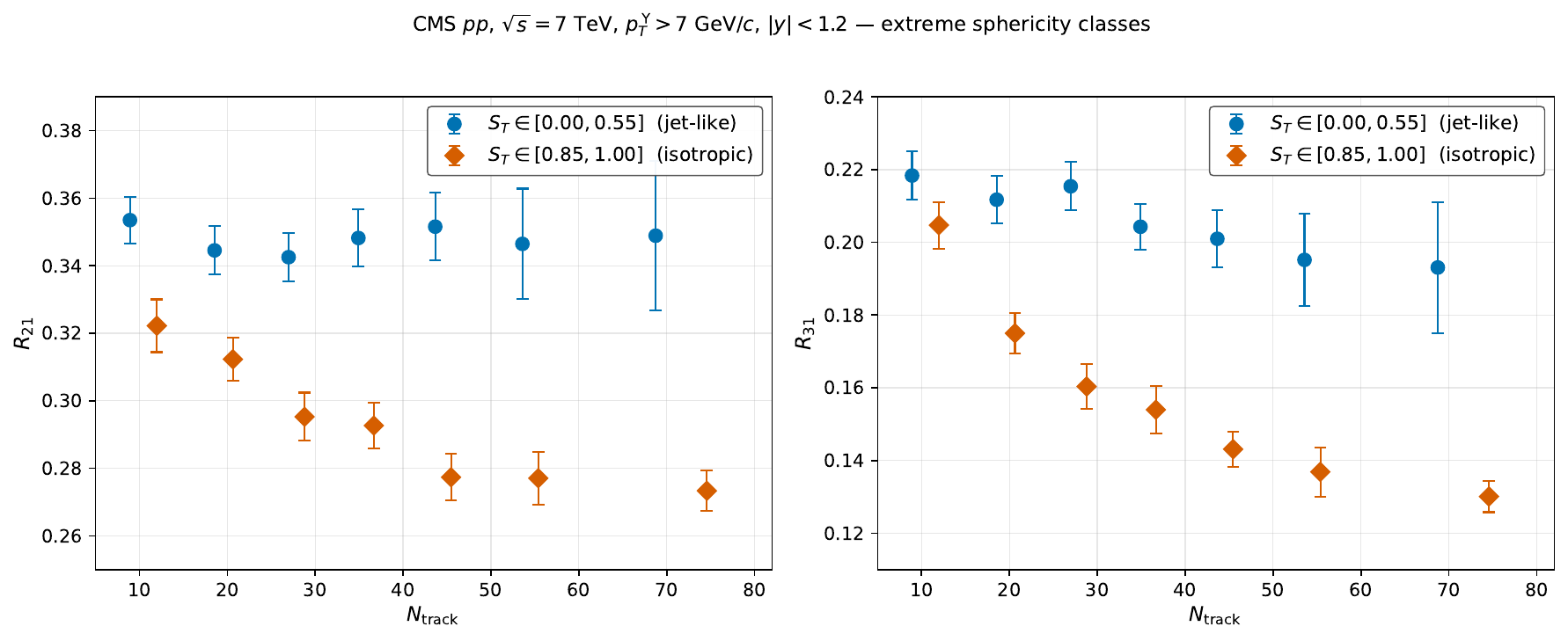}
\caption{Sphericity test: extreme classes.
Left: $\Rtwo$.
Right: $\Rthree$.
Blue circles: jet-like $\ST\in[0.00,0.55]$.
Red diamonds: isotropic $\ST\in[0.85,1.00]$.
Own plot from HEPData of Ref.~\cite{CMS2020},
$\pT^\Upsilon>7$\,GeV/$c$, $|y|<1.2$.
In jet-like events both $\Rtwo$ and $\Rthree$ are essentially flat in
$\Ntrack$.
In isotropic events both ratios decrease monotonically and strongly.
The two populations are selected to have the \emph{same} $\Ntrack$ but
different topology; the fact that the suppression depends on topology
and not just on $\Ntrack$ falsifies every $\Ntrack$-only mechanism.}
\label{fig:sphericity}
\end{figure}
The two extreme classes show a striking contrast.
In jet-like events ($\ST<0.55$) the slope of $\Rtwo$ versus $\Ntrack$
is compatible with zero ($0.2\sigma$) and that of $\Rthree$ is marginal
($2.4\sigma$): the suppression is essentially absent.
In isotropic events ($\ST>0.85$) the slope of $\Rtwo$ differs from
zero by $6.0\sigma$ and that of $\Rthree$ by $10.0\sigma$.
The larger significance for $\Rthree$ reflects both its steeper slope
(the more loosely bound state is more strongly suppressed) and the
smaller relative uncertainties on $\Rthree$ in the high-$\Ntrack$
isotropic bins.
Significance values are computed here from weighted linear fits to
the seven $\Ntrack$ bins of the public HEPData tables of
Ref.~\cite{CMS2020}; the null hypothesis is a slope of zero.
The two populations have the \emph{same} $\Ntrack$ but different
topology; the suppression therefore depends on topology, not on
multiplicity alone.

\paragraph{Consequence.}
Every mechanism parametrised by $\Ntrack$ alone --- hadronic global
CIM, CGC in its standard-form $Q_s^2(\Ntrack)$ reading, pure MPI ---
is in direct tension with the sphericity-resolved data at fixed
$\Ntrack$.
Together with Sec.~\ref{sec:cone} this produces a symmetric closure:
the local reading of the CIM is in tension with the cone test;
the global reading with the sphericity test.
No continuous parameter interpolates between the two without re-opening
one of the failures.
This is the \emph{scissors constraint}: no simple reinterpretation of
the comover density can simultaneously account for cone-independence
\emph{and} sphericity-dependence.
We frame this as a joint constraint on \emph{the naturally expected}
forms of the CIM rather than as an exclusion of the model class as a
whole: a comover-like mechanism that acquires an explicit dependence
on event topology, or that couples to partonic rather than hadronic
degrees of freedom, is not ruled out by this argument alone and could
in principle survive the constraint.

Keeping the three levels separate:
\emph{Data}: at the highest accessible $\Ntrack$ the isotropic class
shows $\Rtwo$ suppressed by ${\sim}15\%$ and $\Rthree$ by
${\sim}25\%$ relative to the jet-like class at the same
$\Ntrack$~\cite{CMS2020}; the effect is stronger for the more loosely
bound state.
\emph{Experimental constraint}: the suppression depends on the
\emph{topology} of the event at fixed multiplicity, not on
multiplicity alone; and the sequential hierarchy --- $\Rthree$
more suppressed than $\Rtwo$ --- is amplified in the isotropic class
relative to the jet-like one.
\emph{Interpretation, with caveats}: the topology dependence is most
naturally read as evidence that the effective medium density, rather
than the track count, is what drives the dissociation; but a
non-trivial differential feed-down pattern between the two event
classes could in principle shift the quantitative picture.

\section{Temporal constraint: LHCb non-prompt $\psipr/\Jpsi$}
\label{sec:nonprompt}
A clean temporal handle is provided by non-prompt quarkonia: states
produced in the weak decay of a $b$ hadron have $c\tau_B\simeq 450$\,$\mu$m,
a proper decay time of $\simeq 1.5$\,ps $\simeq 4.5\times 10^5$\,fm/$c$.
The typical lifetime of any dense system formed in a $pp$ collision is
$\mathcal{O}(1)$\,fm/$c$: the ratio of the two timescales is $\simeq
10^{9}$.
A non-prompt charmonium is therefore born \emph{outside} the initial
dense region and, by causality, cannot be affected by a short-lived
early medium.

\begin{figure}[H]
\centering
\includegraphics[width=0.85\textwidth]{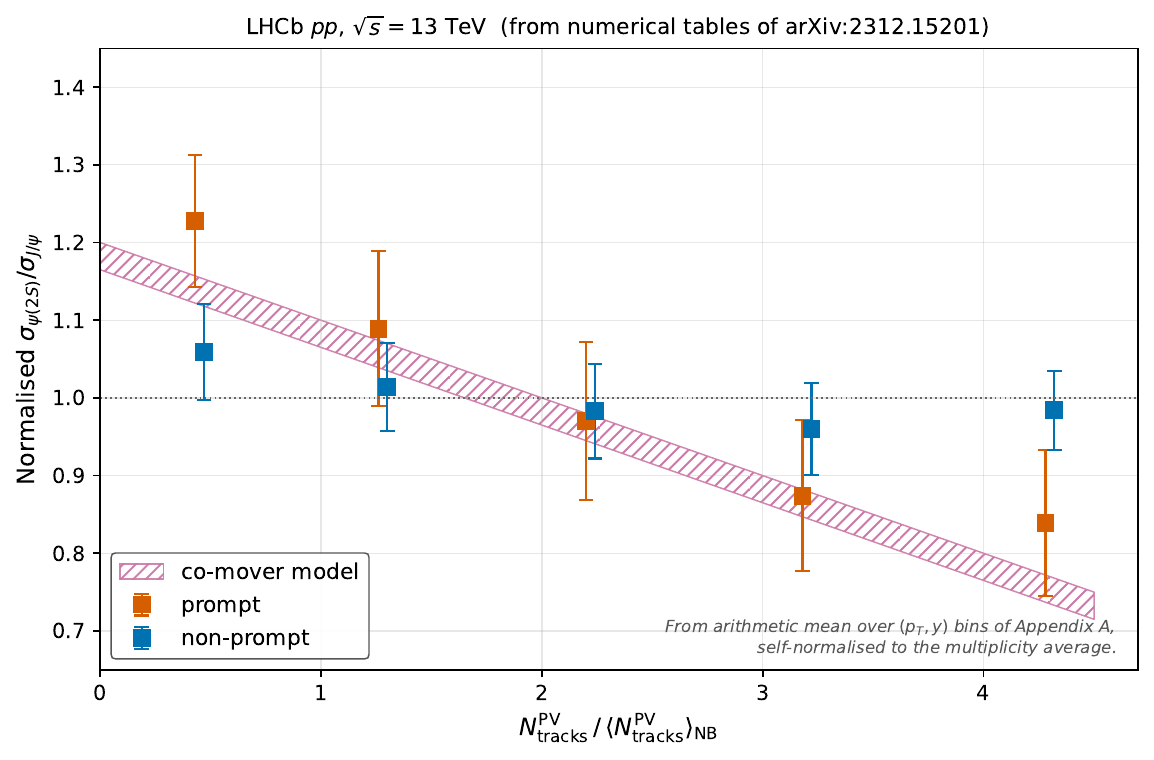}
\caption{Normalised $\sigma_{\psipr}/\sigma_{\Jpsi}$ as a function of
the non-dimensionalised charged-track multiplicity,
$N^{\mathrm{PV}}_{\mathrm{tracks}}/\langle N^{\mathrm{PV}}_{\mathrm{tracks}}\rangle_{\mathrm{NB}}$,
separately for prompt (orange) and non-prompt (blue) production.
Values computed from the numerical Tables 4--8 of Appendix A of
Ref.~\cite{LHCb2024} (arithmetic mean of the double-differential
ratios over the 15 cells in $(p_T, y)$ of each multiplicity bin,
then self-normalised to the mean across the five multiplicity bins).
The co-mover model prediction~\cite{LHCb2024} is shown as the hatched
band.
Error bars are approximate (statistical propagation of the
per-cell uncertainties assuming uncorrelated errors); systematic
uncertainties are not shown.
The prompt ratio decreases monotonically and is well described by
the co-mover model band except in the low-multiplicity region.
The non-prompt ratio is flat within uncertainties across the full
range.}
\label{fig:nonprompt}
\end{figure}

The LHCb 13\,TeV result~\cite{LHCb2024} is structurally clean: the same
detector, same event, same pair of states, the \emph{only} difference
between the two datasets is the decay time.
The prompt component falls; the non-prompt does not.
Any mechanism based on late-stage hadronic rescattering accumulated over
macroscopic distances is inconsistent with this flatness.
The mechanism must act \emph{early}, before displaced $b$-decays occur
and before any late hadronic phase develops.
Because non-prompt charmonia are displaced from the primary vertex by
$\sim 450$\,$\mu$m, they can only be affected by a long-lived medium,
so their flatness forces the prompt-suppression mechanism to act at
early proper times, on the pre-resonance $\ccbar$ (and by extension
$\bbbar$) stage.

A complementary reading is obtained by resolving the prompt and
non-prompt ratios in slices of the charmonium transverse momentum
$p_T^\psi$.
\begin{figure}[H]
\centering
\includegraphics[width=\textwidth]{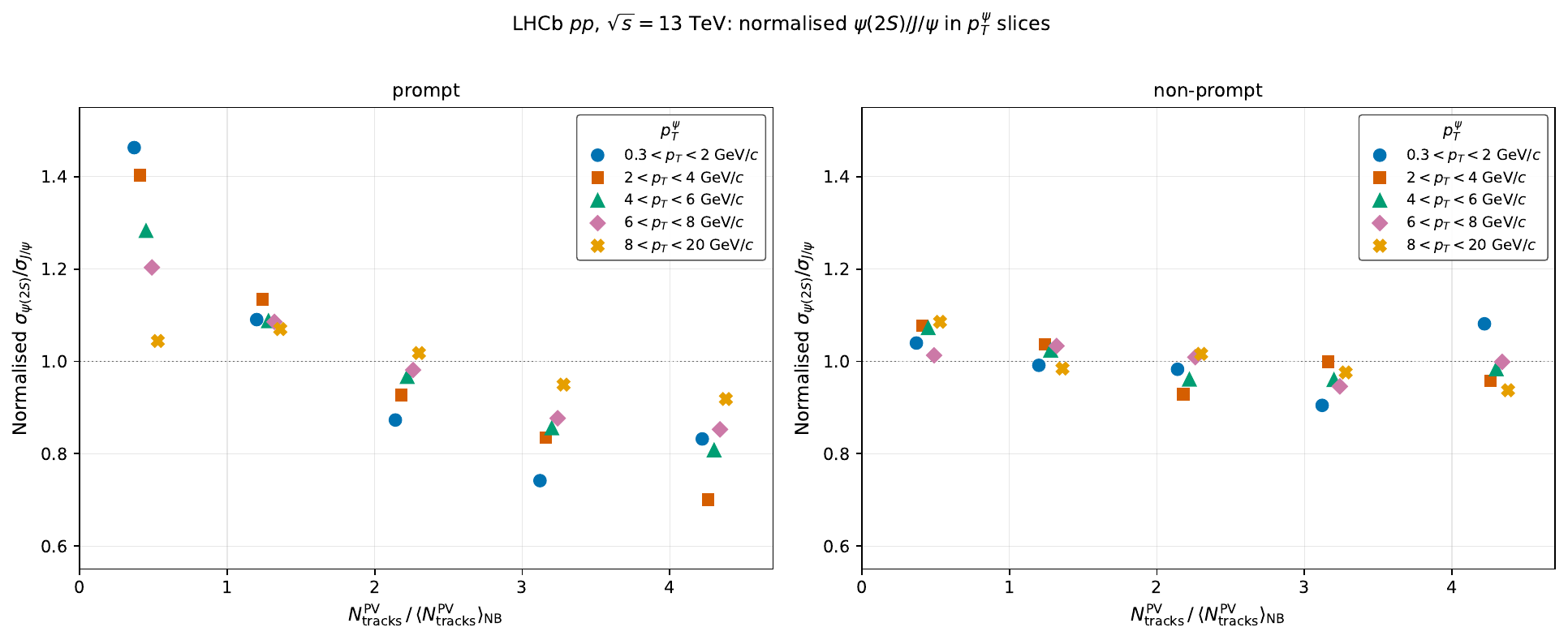}
\caption{LHCb $pp$ data at $\sqrts=13$\,TeV: normalised
$\psipr/\Jpsi$ cross-section ratio as a function of the
non-dimensionalised multiplicity, in five exclusive $p_T^\psi$ slices
($0.3<p_T<2$, $2<p_T<4$, $4<p_T<6$, $6<p_T<8$, $8<p_T<20$\,GeV/$c$),
separately for prompt (left) and non-prompt (right) production.
Own plot from the numerical Tables 4--8 of Appendix A of
Ref.~\cite{LHCb2024}; the self-normalisation is performed slice by
slice to the average of the five multiplicity bins of the same slice.
For prompt production the low-$p_T$ slices show the strongest
multiplicity dependence, and the high-$p_T$ slice ($8<p_T<20$,
yellow crosses) becomes compatible with flat within uncertainty.
For non-prompt production no $\pT$ slice shows a significant
multiplicity trend.
This is the $\psipr/\Jpsi$ analogue of the CMS $\Upsilon$ evidence in
Fig.~\ref{fig:ptslices}.}
\label{fig:lhcb_ptslices}
\end{figure}
Two observations parallel the CMS picture of Sec.~\ref{sec:basic}:
(i) among prompt charmonia, the suppression weakens as $p_T^\psi$
grows, so that the highest-$p_T$ slice is compatible with no
multiplicity dependence within uncertainty --- the kinematic-escape
pattern already observed for $\Upsilon(n\mathrm{S})$;
(ii) for non-prompt charmonia the ratio is flat irrespective of
$p_T^\psi$, so the flatness reported in Fig.~\ref{fig:nonprompt} is
not the effect of integrating over $p_T$ but holds slice by slice.
Together, Figs.~\ref{fig:nonprompt} and~\ref{fig:lhcb_ptslices} show
that the $\pT^\psi$-dependence and the prompt/non-prompt dichotomy are
two \emph{independent} features of the LHCb data, each carrying a
separate physical constraint: a finite-size medium (from the
$\pT^\psi$ ordering) acting at early proper times (from the
prompt-only trend).

One caveat is that the non-prompt argument is measured in charmonium,
while the sequential suppression pattern of Sections~\ref{sec:basic}
to~\ref{sec:sphericity} concerns bottomonium.
The logical content we use here is not a flavour-specific statement but
a \emph{temporal} one: whatever mechanism produces the prompt-only
multiplicity dependence must act at proper times short compared to the
displacement of $b$-hadron decays, i.e.\ at the pre-resonance stage.
This kind of temporal bound is model-independent and applies equally
to any quarkonium, charmonium or bottomonium, since both are formed
within $\mathcal{O}(1)$\,fm/$c$ of the primary vertex.
We therefore take the non-prompt flatness as a general constraint on
the \emph{kind} of mechanism at work rather than as direct evidence on
$\ups{n}$ specifically.

\section{Review of hadronic and string-based frameworks}
\label{sec:models}
We confront, one by one, the frameworks that have been proposed to
explain the suppression in $pp$ with the four independent filters
(cone, azimuth, sphericity, $\pT$-escape) plus the non-prompt flatness.

\paragraph{Local hadronic CIM.}
Dissociation by nearby hadrons; the model predicts \emph{more}
suppression in the dense-cone class than in the empty-cone class.
The CMS data show the two classes to be indistinguishable within
uncertainty.
Verdict: \textbf{in direct tension with the cone test}; the model
in its local-density reading does not reproduce the sign of the
predicted effect.

\paragraph{Global hadronic CIM ($\rho_{\mathrm{com}}\propto\Ntrack$).}
Prediction: cone- and azimuth-independent (satisfied), but
\emph{shape-independent} at fixed $\Ntrack$.
Data: $\Rtwo$ and $\Rthree$ both strongly shape-dependent
(Fig.~\ref{fig:sphericity}).
Verdict: \textbf{incompatible with the sphericity-resolved data in its
$\rho_{\mathrm{com}}\propto\Ntrack$ form}.

\paragraph{Pure MPI (PYTHIA 8, no extra physics)~\cite{Sjostrand2015}.}
Multi-parton interactions generate $\Ntrack$ but do not interact with
the produced $\Upsilon$.
Prediction: $R_{n1}$ flat in $\Ntrack$.
Data: monotonic decrease, enhanced in isotropic events.
Verdict: \textbf{outside the model's predictive scope}; PYTHIA\,8 with
only MPI is not designed to reproduce a multiplicity-dependent
$\Upsilon(n\mathrm{S})$ hierarchy.

\paragraph{Colour reconnection / rope hadronisation~\cite{Bierlich2015}.}
Overlapping strings from many MPI form ropes with enhanced effective
tension $\kappa_{\mathrm{eff}}>\kappa_0$, preferentially suppressing
loosely-bound states.
Cone: OK (string overlaps set at production time, not driven by which
cone the fragments end up in).
Azimuth: OK (strings approximately uniform in $\phi$).
Sphericity: the direction of the effect is correct (isotropic
events have more MPI and hence larger rope tension), but no published
PYTHIA+rope simulation reproduces, at fixed parameters, the observed
gap between the two extreme classes: $\Rtwo$ compatible with flat
($0.2\sigma$) and $\Rthree$ marginal ($2.4\sigma$) in jet-like events,
against $6.0\sigma$ and $10.0\sigma$ respectively in isotropic events
at the same $\Ntrack$~\cite{CMS2020}.
$\pT$-escape: the rope tension does not depend on $\pT^\Upsilon$, so
the model has no native escape mechanism; the LHCb low-$\pT$ enhancement
of suppression is not naturally reproduced.
Verdict: \textbf{qualitatively consistent with cone and azimuth,
quantitatively unvalidated on sphericity, and structurally missing a
native $\pT$-escape}; not ruled out as a partial contribution but not
a complete explanation on present evidence.

\paragraph{Colour Glass Condensate (initial-state saturation)~\cite{Ma2015}.}
Quarkonium dissociated by a high-density gluon field with saturation
scale $Q_s^2$, set before the collision.
Cone: OK. Azimuth: OK.
Sphericity: a CGC calibrated by the overall initial-state density is
$\Ntrack$-driven, and therefore predicts identical suppression in
jet-like and isotropic events at fixed $\Ntrack$, contrary to the
data.
$\pT$-escape: no finite-size geometry, no native escape mechanism.
Verdict: \textbf{in tension with the sphericity-resolved data and
structurally missing a $\pT$-escape structure} in its standard form.

\paragraph{Two-component underlying-event parametrisations~\cite{Trainor2008}.}
A recurring alternative reading of the high-multiplicity $pp$ data is
Trainor's two-component model (TCM), in which every spectrum is
decomposed into a ``soft'' component (scaling with the number of
participant partons) and a ``hard'' component (scaling with the number
of hard MPI), without invoking a medium.
Applied to quarkonium, TCM re-expresses $R_{n1}(\Ntrack)$ as an
interpolation between a soft plateau and a hard-dominated tail: a
suppression emerges from the fact that the soft component carries a
different effective $R_{n1}$ than the hard component, and their
relative weight shifts with $\Ntrack$.
Tuned per observable, the model can be made to track either $\Rtwo$ or
$\Rthree$ individually.
The problem is that the differential structure of the CMS data fixes
the $\Ntrack$-dependence of the weights in a way that is not
self-consistent across cuts: at fixed $\Ntrack$, jet-like and isotropic
events have very different soft/hard decompositions
(jet-like: hard-component dominated; isotropic: soft-component
dominated) so the TCM prediction is \emph{different} for the two sphericity
classes, but the magnitude of the split does not reproduce the
sphericity-resolved data; likewise, cone-empty and cone-dense events
cannot be selected to simultaneously match the observed near-equality
in the cone test and the observed strong gap in the sphericity test.
In practice no public TCM fit has reproduced the full multi-differential
CMS set at fixed parameters.
Verdict: \textbf{no demonstrated predictive content once the full
multi-differential set is imposed}; the model retains flexibility but
has yet to match the combined constraints.

\begin{table}[H]
\centering
\caption{Verdict matrix.
OK = the framework accounts for the feature naturally;
Tension = structural difficulty in the framework's standard/published form;
Tunable = can be accommodated with parameter adjustments.
No entry is meant as a statement that \emph{every} variant of a framework
is excluded; it reflects the status of the \emph{considered} published
formulations.}
\label{tab:verdict}
\begin{tabular}{lccccc}
\toprule
Framework & Seq.\ suppr. & Cone & Sphericity & $\pT$-escape & Non-prompt flat \\
\midrule
Local hadronic CIM          & Tension    & Tension   & Tension   & Tunable  & OK \\
Global hadronic CIM         & Tension    & OK        & Tension   & Tunable  & OK \\
Pure MPI (PYTHIA)           & Tension    & OK        & Tension   & Tension  & OK \\
CR / Rope hadronisation     & Tunable    & OK        & Tension   & Tunable  & OK \\
CGC (initial state)         & Tunable    & OK        & Tension   & Tension  & OK \\
Two-component UE            & Tension    & Tension   & Tension   & Tension  & OK \\
\midrule
Globally correlated medium  & OK         & OK        & OK        & OK       & OK \\
\bottomrule
\end{tabular}
\end{table}
None of the considered frameworks naturally and simultaneously accounts
for the combined set of constraints in its published form.
The entry in the last row is not a specific model but a description of
the minimal structural features a surviving framework must have: it must
act early (non-prompt flat), globally (sphericity), non-locally around
the $\Upsilon$ (cone), and must carry a $\pT^\Upsilon$-dependent
escape probability.
The natural physical realisation of all four features simultaneously
is an early, globally correlated medium with partonic degrees of
freedom.

\section{Multi-observable context: strangeness, ridge, partonic flow}
\label{sec:context}
The onset of $\Upsilon$ suppression in the CMS data is at
$\Ntrack\approx 10$--$15$, corresponding to $\dNdeta\approx 4$--$6$
(using the conversion $\Ntrack\simeq 2.83\,\dNdeta$).
Three independent collective-like phenomena in the same collision
system share the same onset window.

\paragraph{(a) Strangeness enhancement.}
ALICE~\cite{ALICE_SE} measured a monotonic enhancement of strange and
multi-strange hadrons (K$_S^0$, $\Lambda$, $\Xi$, $\Omega$) relative to
charged pions, rising by factors $1.5$--$2.5$ from low to high
multiplicity at $\sqrts=7$\,TeV.
The onset falls at $\dNdeta\approx 4$--$8$.
Strangeness enhancement is the canonical QGP signature from heavy-ion
physics.
At $\sqrts=5.02$\,TeV~\cite{ALICE_SE5TeV} the high-multiplicity tail of
the $pp$ data is better described by a generator with a QGP
core (EPOS LHC~\cite{Werner2014}) than by pure string fragmentation
(PYTHIA 8).
The broader context of ``small-system collectivity'' --- ridge,
partonic-like $v_n$ harmonics, strangeness, radial flow --- as it
evolved in the decade preceding this paper is reviewed in
Ref.~\cite{NagleZajc}.

\paragraph{(b) Ridge: long-range near-side correlations.}
Two-particle correlations in $\Delta\eta$--$\Dphi$ in high-multiplicity
$pp$ at the LHC show a near-side ridge at $\Dphi\simeq 0$ extending
over several units of $\Delta\eta$~\cite{CMS_ridge}.
The ridge is a hallmark of early-time, long-range correlations in the
initial-state gluon field, which persist through the evolution.
The long-$\Delta\eta$ structure of the ridge is causally compatible only
with an interaction at very short times, matching the non-prompt
constraint of Sec.~\ref{sec:nonprompt}.

\paragraph{(c) Partonic flow (baryon--meson $v_2$ grouping).}
ALICE 2024~\cite{ALICE_v2} reports the first observation, with
$\sim 5\sigma$ significance, of baryon--meson $v_2$ grouping in
high-multiplicity $pp$ at $\sqrts=13$\,TeV.
This is the mass ordering at low $\pT$ giving way to a baryon--meson
ordering at intermediate $\pT$: the canonical signature of
\emph{partonic} collective flow with number-of-constituent-quark
scaling.
It requires the top $0.07\%$ of events in V0M multiplicity,
$\dNdeta\approx 26$ at midrapidity, and is \emph{absent} in the
low-multiplicity class.

\paragraph{Hierarchy of onsets.}
\begin{center}
\begin{tabular}{lc}
\toprule
Observable & Onset ($\dNdeta$) \\
\midrule
$\Upsilon(2$S$,3$S$)$ sequential suppression & $\approx 4$--$6$ \\
Strangeness enhancement                       & $\approx 4$--$8$ \\
Baryon--meson $v_2$ grouping (partonic flow)  & $\gtrsim 17$--$26$ \\
\bottomrule
\end{tabular}
\end{center}
The $\Upsilon(n\mathrm{S})$ suppression begins before the explicit
partonic-flow signal becomes measurable.
This is not paradoxical: the loosely bound excited states have binding
energies of a few hundred MeV ($\sim 200$\,MeV for $\ups{3}$) and
therefore are sensitive to a density regime in which partonic flow,
set by the number of constituents participating coherently, is still
below the detection threshold.
The quarkonium is the earliest probe of the transition; partonic $v_2$
requires more developed collectivity before becoming measurable.

\section{Onset multiplicity window: consistency with the Campanini \& Ferri EOS}
\label{sec:eos}
The four differential constraints discussed in
Secs.~\ref{sec:cone}--\ref{sec:nonprompt} all become active in the same
multiplicity window: $\dNdeta\approx 4$--$6$ at midrapidity.
This is the onset of the suppression signal, of the strangeness
enhancement (Sec.~\ref{sec:context}), and of the ridge; the
baryon--meson $v_2$ grouping switches on at $\dNdeta\gtrsim 17$.

On independent grounds, the equation-of-state analysis of
Campanini \& Ferri~\cite{Campanini2011}, performed on inclusive $pp$
and $p\bar p$ multiplicity distributions across ISR-to-LHC energies,
identifies a soft-to-hard transition band at $\dNdeta\approx 6$--$24$.
This band is determined from the EOS proxy $\sigma_s/\langle p_T\rangle^3$
without reference to any quarkonium observable.
The coincidence of the two onset windows is noted here as a consistency
observation.
It suggests that the density regime in which the multi-differential
CMS and LHCb constraints point toward an early, globally correlated
medium is also the regime in which the simplest hadronic description
reaches the limit of its applicability.
No energy-density estimate is required to reach this conclusion: the
common variable throughout is $\dNdeta$, a directly measured,
model-independent quantity.

\section{Why jet quenching is \emph{not} visible --- and why that is
consistent}
\label{sec:nojet}
A legitimate question is: if a partonic medium is present in
high-multiplicity $pp$, why is jet quenching not observed?
Two independent, quantitative reasons make the absence of jet
quenching compatible with, and indeed expected from, the present
picture.

\paragraph{(i) Path-length scaling.}
The path-length dependence of parton energy loss depends on the
dominant mechanism.
For radiative (gluon-radiation) loss in the LPM regime,
\begin{equation}
\Delta E_{\mathrm{rad}} \propto \hat q\,L^2,
\end{equation}
with $\hat q$ the transport coefficient and $L$ the path length.
For purely collisional (elastic) loss the scaling is linear,
$\Delta E_{\mathrm{coll}} \propto \hat q'\,L$.
In central PbPb the medium extends over $L_{\mathrm{PbPb}}\sim 5$--$6$\,fm;
in $pp$ the droplet extends over $L_{pp}\sim 1$--$2$\,fm.
The two parametrisations bracket the uncertainty:
\begin{equation}
\frac{\Delta E_{pp}}{\Delta E_{\mathrm{PbPb}}}
\sim
\begin{cases}
\left(\dfrac{L_{pp}}{L_{\mathrm{PbPb}}}\right)^2 \sim \tfrac{1}{25}\text{--}\tfrac{1}{36}
& \text{(radiative)}\\[8pt]
\dfrac{L_{pp}}{L_{\mathrm{PbPb}}} \sim \tfrac{1}{4}\text{--}\tfrac{1}{5}
& \text{(collisional)}
\end{cases}
\end{equation}
An effect that is tens of GeV in central PbPb becomes at most a few GeV
even in the most favourable (linear) scaling, and more likely a few
hundred MeV in the radiative regime.
In both cases it falls below the current sensitivity of jet measurements
in $pp$, which require $\mathcal{O}(\text{GeV})$-scale modifications to
be resolved above the hadronic calorimetry threshold.
The effect is therefore below the detection floor, not absent, under
any plausible scaling assumption.

\paragraph{(ii) Quarkonium sensitivity scale.}
A few hundred MeV are invisible to jet quenching but
\emph{decisive for the $\ups{3}$}, whose binding energy is
$\sim 200$\,MeV.
A medium whose jet-quenching footprint is negligible is nonetheless
capable of dissolving loosely bound quarkonia efficiently.
The sensitivity of sequential $\Upsilon(n\mathrm{S})$ suppression to
the density range
\begin{center}
\emph{above} the deconfinement crossover, \emph{below} the
jet-quenching detection threshold,
\end{center}
is structural.
The experimental facts --- strong sequential $\Upsilon$ suppression,
strangeness enhancement, ridge, partonic $v_2$, no jet quenching --- are
all simultaneously consistent with a density regime in which
$\Upsilon(n\mathrm{S})$ states, and not high-$\pT$ jets, are the
uniquely sensitive probes.

\section{Interpretation and closure}
\label{sec:closure}
Four independent differential filters (cone, azimuth, sphericity,
$\pT$-escape) and one temporal filter (non-prompt flat) act together on
the data.
Each filter places a different constraint on the admissible mechanisms.
In their published forms, the hadronic and string-based frameworks
considered in Sec.~\ref{sec:models} fail at least one of these
constraints without room for an obvious, parameter-free repair.
On the evidence presented, the minimal set of structural features a
surviving framework must carry is that of an early, globally correlated
medium consistent with partonic degrees of freedom.
The qualitative ingredients are:
\begin{enumerate}
\item Finite volume: core--corona with $R_{\mathrm{HBT}}\approx 1.8$\,fm,
a survival floor for corona quarkonia, and a central region that
dissolves the excited states.
\item Momentum-dependent escape: crossing time
$\tau_{\mathrm{cross}}\sim R/(\beta\gamma)$ becomes shorter than the
medium lifetime at high $\pT$, generating the low-/high-$\pT$ asymmetry
observed by LHCb.
\item Early action: the non-prompt flatness fixes the mechanism to act
on the pre-resonance $\bbbar$ (or $\ccbar$) stage, compatible with
Debye-like colour screening.
\item Onset window coincidence: the multiplicity range in which all
four differential constraints become active ($\dNdeta\approx 4$--$6$)
coincides with the soft-to-hard transition band identified by
Campanini \& Ferri~\cite{Campanini2011} from inclusive $pp$ and
$p\bar p$ data across ISR-to-LHC energies, without reference to any
quarkonium observable.
\end{enumerate}
The onset co-occurrence with strangeness enhancement and ridge, and the
subsequent turning on of partonic $v_2$ at higher multiplicity, form a
coherent multi-observable picture that is difficult to accommodate as a
chance superposition of unrelated effects.

\paragraph{What this analysis does and does not claim.}
This is a constraint paper.
What it establishes, to the extent that the public CMS and LHCb datasets
allow, is that the multi-differential structure of the measured
$\Upsilon(n\mathrm{S})$ and $\psipr/\Jpsi$ patterns places strong
joint requirements on any candidate mechanism, and that the hadronic
and string-based frameworks we have been able to examine do not
naturally satisfy those requirements simultaneously.
The analysis should be read as a map of what any future framework must
reproduce, and as an experimental case --- not a formal demonstration ---
that the most economical reading of the combined evidence involves an
early, globally correlated, partonic-like medium.

\section*{Data availability and source attribution}
CMS 7\,TeV data: HEPData \texttt{ins1805867} (doi:\href{https://doi.org/10.17182/hepdata.95684.v1}{10.17182/hepdata.95684.v1});
LHCb 13\,TeV $\psipr/\Jpsi$ data: JHEP \textbf{05} (2024) 243, Fig.~2
(CC BY 4.0, reproduced visually in Fig.~\ref{fig:nonprompt}).
All numerical values in this paper come from the public datasets cited;
no new measurement is claimed.

\section*{Use of AI assistance}
The author used Claude (Anthropic, claude-sonnet-4-6) for assistance
with LaTeX editing, numerical cross-checks, and formal writing tasks
during the preparation of this manuscript.
All scientific ideas, analyses, interpretations, and conclusions are
the author's own and have been critically reviewed by the author.



\begin{thebibliography}{99}
\bibitem{CMS2020}
CMS Collaboration,
\emph{Investigation into the event-activity dependence of $\Upsilon(n\mathrm{S})$
relative production in proton--proton collisions at $\sqrts=7$\,TeV},
JHEP \textbf{11} (2020) 001 [arXiv:2007.04277].

\bibitem{LHCb2024}
LHCb Collaboration,
\emph{Multiplicity dependence of $\sigma_{\psi(2S)}/\sigma_{J/\psi}$
in $pp$ collisions at $\sqrts=13$\,TeV},
JHEP \textbf{05} (2024) 243 [arXiv:2312.15201].

\bibitem{LHCb2025}
LHCb Collaboration,
\emph{Measurement of multiplicity-dependent $\Upsilon(n\mathrm{S})$
production ratios in $pp$ collisions at $\sqrts=13$\,TeV},
LHCb-PAPER-2025.

\bibitem{MatsuiSatz1986}
T.~Matsui and H.~Satz,
\emph{$\Jpsi$ suppression by quark-gluon plasma formation},
Phys.\ Lett.\ B \textbf{178} (1986) 416.

\bibitem{Satz2006}
H.~Satz, \emph{Colour deconfinement and quarkonium binding},
J.~Phys.\ G \textbf{32} (2006) R25 [hep-ph/0512217].

\bibitem{Armesto1998}
N.~Armesto and A.~Capella,
\emph{A quantitative model for $\Jpsi$ suppression in nuclear collisions},
Phys.\ Lett.\ B \textbf{430} (1998) 23.

\bibitem{Gavin1990}
S.~Gavin and R.~Vogt,
\emph{Charmonium suppression by Comover scattering in Pb+Pb collisions},
Phys.\ Rev.\ Lett.\ \textbf{78} (1997) 1006 [hep-ph/9606460].

\bibitem{Ferreiro2018}
E.~G.~Ferreiro and J.-P.~Lansberg,
\emph{Is bottomonium suppression in proton--nucleus and nucleus--nucleus
collisions at LHC energies due to the same effects?},
JHEP \textbf{10} (2018) 094 [arXiv:1804.04474].

\bibitem{Ma2015}
Y.-Q.~Ma and R.~Venugopalan,
\emph{Comprehensive description of $\Jpsi$ production in
proton--proton collisions at collider energies},
Phys.\ Rev.\ Lett.\ \textbf{113} (2014) 192301 [arXiv:1408.4075].

\bibitem{Bierlich2015}
C.~Bierlich, G.~Gustafson, L.~L\"onnblad and A.~Tarasov,
\emph{Effects of overlapping strings in $pp$ collisions},
JHEP \textbf{03} (2015) 148 [arXiv:1412.6259].

\bibitem{Sjostrand2015}
T.~Sj\"ostrand et al.,
\emph{An introduction to PYTHIA 8.2},
Comput.\ Phys.\ Commun.\ \textbf{191} (2015) 159 [arXiv:1410.3012].

\bibitem{Trainor2008}
T.~A.~Trainor,
\emph{A two-component model for the transverse-momentum spectra from
$pp$, $p\bar p$ and $AA$ collisions},
Int.\ J.\ Mod.\ Phys.\ E \textbf{17} (2008) 1499 [arXiv:0710.4504].

\bibitem{ALICE_SE}
ALICE Collaboration,
\emph{Enhanced production of multi-strange hadrons in
high-multiplicity $pp$ collisions},
Nature Phys.\ \textbf{13} (2017) 535 [arXiv:1606.07424].

\bibitem{ALICE_SE5TeV}
ALICE Collaboration,
\emph{Strangeness enhancement at its extremes: multiple (multi-)strange
hadron production in $pp$ collisions at $\sqrts=5.02$\,TeV},
arXiv:2511.10413 (2025).

\bibitem{ALICE_v2}
ALICE Collaboration,
\emph{Observation of partonic flow in proton--proton and
proton--nucleus collisions},
Nature Commun.\ \textbf{17} (2026) 2585 [arXiv:2411.09323].

\bibitem{ALICE_spectra}
ALICE Collaboration,
\emph{Multiplicity dependence of light-flavour hadron production in
$pp$ collisions at $\sqrts=7$\,TeV},
Eur.\ Phys.\ J.\ C \textbf{79} (2019) 857.

\bibitem{CMS_ridge}
CMS Collaboration,
\emph{Observation of long-range near-side angular correlations in
proton--proton collisions at the LHC},
JHEP \textbf{09} (2010) 091 [arXiv:1009.4122].

\bibitem{Campanini2011}
R.~Campanini and G.~Ferri,
\emph{Experimental equation of state in proton--proton and
proton--antiproton collisions and phase transition to quark
gluon plasma}, Phys.\ Lett.\ B \textbf{703} (2011) 237
[arXiv:1106.2008].
\bibitem{Werner2014}
K.~Werner, B.~Guiot, Iu.~Karpenko and T.~Pierog,
\emph{Analysing radial flow features in $p$--Pb and $pp$ collisions at
several TeV by studying identified-particle production with the event
generator EPOS3},
Phys.\ Rev.\ C \textbf{89} (2014) 064903 [arXiv:1312.1233].

\bibitem{Esposito2021}
A.~Esposito, E.~G.~Ferreiro, A.~Pilloni, A.~D.~Polosa and
C.~A.~Salgado,
\emph{The nature of $X(3872)$ from high-multiplicity $pp$ collisions},
Eur.\ Phys.\ J.\ C \textbf{81} (2021) 669 [arXiv:2006.15044].

\bibitem{NagleZajc}
J.~L.~Nagle and W.~A.~Zajc,
\emph{Small system collectivity in relativistic hadronic and nuclear
collisions},
Annu.\ Rev.\ Nucl.\ Part.\ Sci.\ \textbf{68} (2018) 211
[arXiv:1801.03477].

\end{thebibliography}
\end{document}